\documentclass[10pt,a4paper]{article}
\usepackage[utf8]{inputenc}
\usepackage{color}
\usepackage{cmap}
\usepackage{amsmath}
\usepackage{amsfonts}
\usepackage{amssymb}
\usepackage{graphicx}
\usepackage{caption}
\usepackage{subcaption}
\usepackage{setspace}
\graphicspath{ {./Images/} }
\graphicspath{ {./Figures/} }

\title{How to Choose the Best CMOS Technology for Your Application}
\author{Dina Reda El-Damak}
\date{December 2022}

\begin{document}

\maketitle

\section{Introduction}

Complementary-metal-oxide-semiconductor (CMOS) is the most widely spread technology for integrated circuits fabrication. Each foundry offers different technology nodes that are characterized by the minimum feature size, which is the smallest dimension that can be fabricated using photo lithography with the required precision. The minimum feature size ranges from 350 nm to 5nm. Each designer has to select the best technology for implementing their designs, and the designer has to specify the foundry, the technology node, and any additional features like the number of metal layers, and MIM capacitor density. In this white paper, we will describe the various trade-offs while selecting the optimal technology node for power management. You will learn 1) Semiconductor products business models 2) fabrication technologies features 3) how to select the optimal technology node for fabrication according to the application, 4) how the cost of a silicon chip scales with the sales of your design.

\section{Semiconductor Product Business Models}

There are different business models for semiconductor products and IC design companies:

\begin{enumerate}

 \item  Category 1: Designing a commercial product that will we be sold to end-customer. Examples of such products are smart phones, and smart-home assistant devices. In this case, the goal is select the process that guarantee the overall product competitiveness depending on the market. For example, the majority of the customers are cost-oriented in the developing countries, while the customers are performance-oriented in the developed countries.
 
 \item Category 2:  Designing a stand-alone integrated circuit that will be packaged and sold independently to manufacturers of commercial electronic products from category 1. For example, designing the processor of the smart phone, or tablet. In this case, the goal is select the optimal process that reduce the cost, and improves the performance relative to competitors.
 
 \item  Category 3: Designing an IP that an IC design from category 2 can integrate in a System-On-a-Chip (SoC). For example, designing standard cells or LDO without packaging. The design has to follow the foundry guidelines in order to ease its integration with various customer's designs. In this situation, the foundry determine the process node.
 
 \item Category 4: Designing an integrated circuit for a a specific SoC. In this case, the owner of the SoC determines the process.
\end{enumerate}

In this white paper, we will focus on the first and second business models, where the designer is involved in the selection of the optimal technology.

\section{CMOS Technologies Features}

\begin{figure}[ht]
\centering
\includegraphics[width=\textwidth]{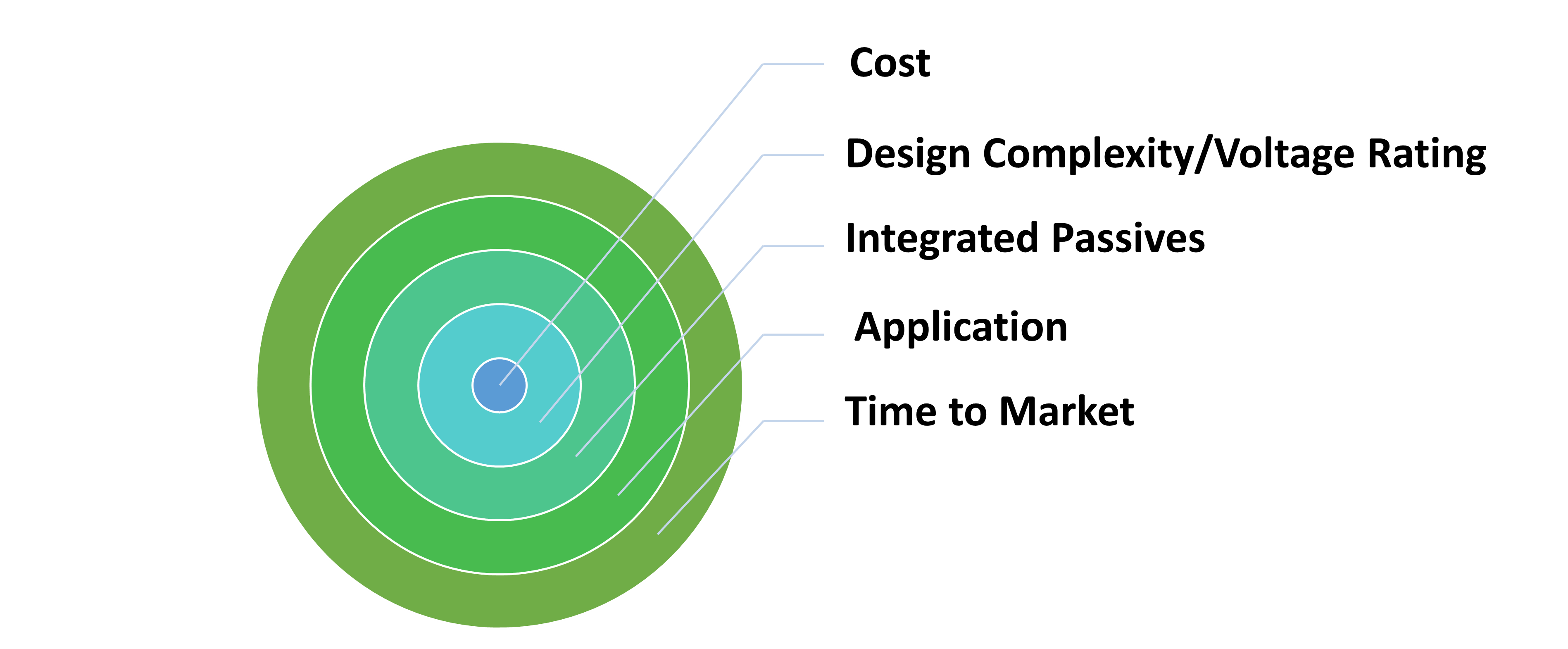}
\caption{Factors affecting the selection of the optimal CMOS Fabrication Technology}
\label{tradeoff}
\end{figure}

CMOS technology selection is the first important step in any integrated circuit project. Tradeoff exists between (1) cost (number of fabrication masks, lithography precision), (2) Design complexity due to voltage rating (3) accessibility to special passives (high density capacitors, integrated magnetics), (4) maximum switching frequency, (5) time to market (number of fabrication slots per year) as shown in figure \ref{tradeoff}. For example, a comparison between TSMC 180nm, 180nm High Voltage, 65nm, and 28nm is shown in \cite{TSMC_comparison}. The technology provider specifies the core(V), I/O (V), the density of the integrated capacitors, the minimum area for fabrication, frequency of fabrication, and the price per mm\textsuperscript{2}. The guidelines for understanding this comparison is listed below.

\begin{itemize}
    \item \textbf{Core Voltage:} The core voltage is the maximum voltage rating of the transistors in the process design kit. The higher the core voltage, the easier to interface with lithium batteries without the need for stacking or complex gate drivers.
    
    \item \textbf{Capacitors Density:} The higher the density of the capacitors, the smaller the area of the switched-capacitor DC-DC converters, and decoupling capacitors on-chip. The frequency of fabrication is another important factor in selection of fabrication process. 
    
    \item \textbf {Minimum Area for Fabrication:} While the foundry price for fabrication is typically listed per millimeter squared. The designer have to be careful that the foundry might also dictates a minimum area for fabrication.
    
    \item \textbf {Frequency of Fabrication:}  If you have a choice, select a process with frequent fabrication opportunities to reduce the risk of missing the foundry deadline.For example, TSMC 180 nm and 65nm Technology nodes have 12 MPW shuttles per year. Figure \ref{GF_TSMC} depicts the the number of MPW shuttles of Global Foundaries (GF) and TSMC. 

\item \textbf {Price of Fabrication:} Figure \ref{GF_TSMC} depicts the price of fabrication of 1 mm squared for various technology nodes from Global Foundaries, and TSMC. The price includes fabrication of the photo mask and a number of silicon chips. The photo mask cost is a fixed cost for the semiconductor devic 

The price of  fabrication of 1 millimeter squared of 180 nm starts from \$ 1100 (which is equilvalent to 21066.76 L.E./mm\textsuperscript{2} for TSMC 180 nm General-Purpose technology according to the US Dollars and Egyptian pound exchange rate on  August 12, 2022). However, the price of 1 millimeter squared using advanced technology nodes like 12nm is 30X higher around \texteuro 31,200 (which is equivalent to  615592.64 L.E./mm\textsuperscript{2} for GF 12nm technology according to the Euros and Egyptian pounds exchange rate on  August 12, 2022) \cite{Europractice}. In addition to controlling the cost by the minimum feature size of the technology, the foundry might request additional fees for supporting High Voltage (HV) Devices, Non-Volatile Memory (NVM), OPTO process, Silicon on Insulator (SOI) as shown in Figure \ref{XFAB}. 

\begin{figure}[ht]
\centering
\includegraphics[width=\textwidth]{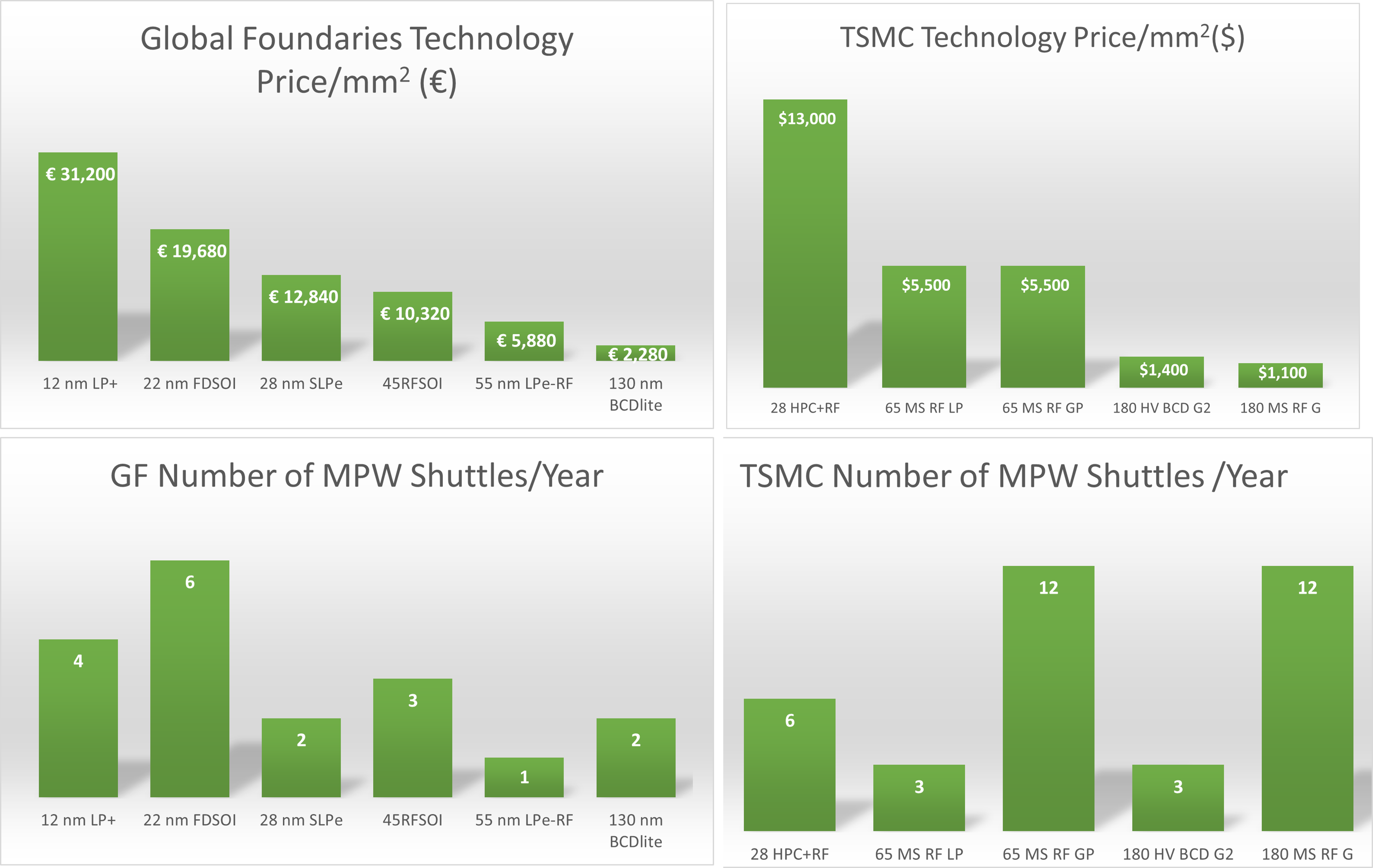}
\caption{Global Foundaries and TSMC Technology Price per millimeter squares and frequency of MPW shuttles [Source:Muse Semiconductor and Europractice \cite{TSMC_comparison,Europractice}]}
\label{GF_TSMC}
\end{figure}

\begin{figure}[ht]
\centering
\includegraphics[width=\textwidth]{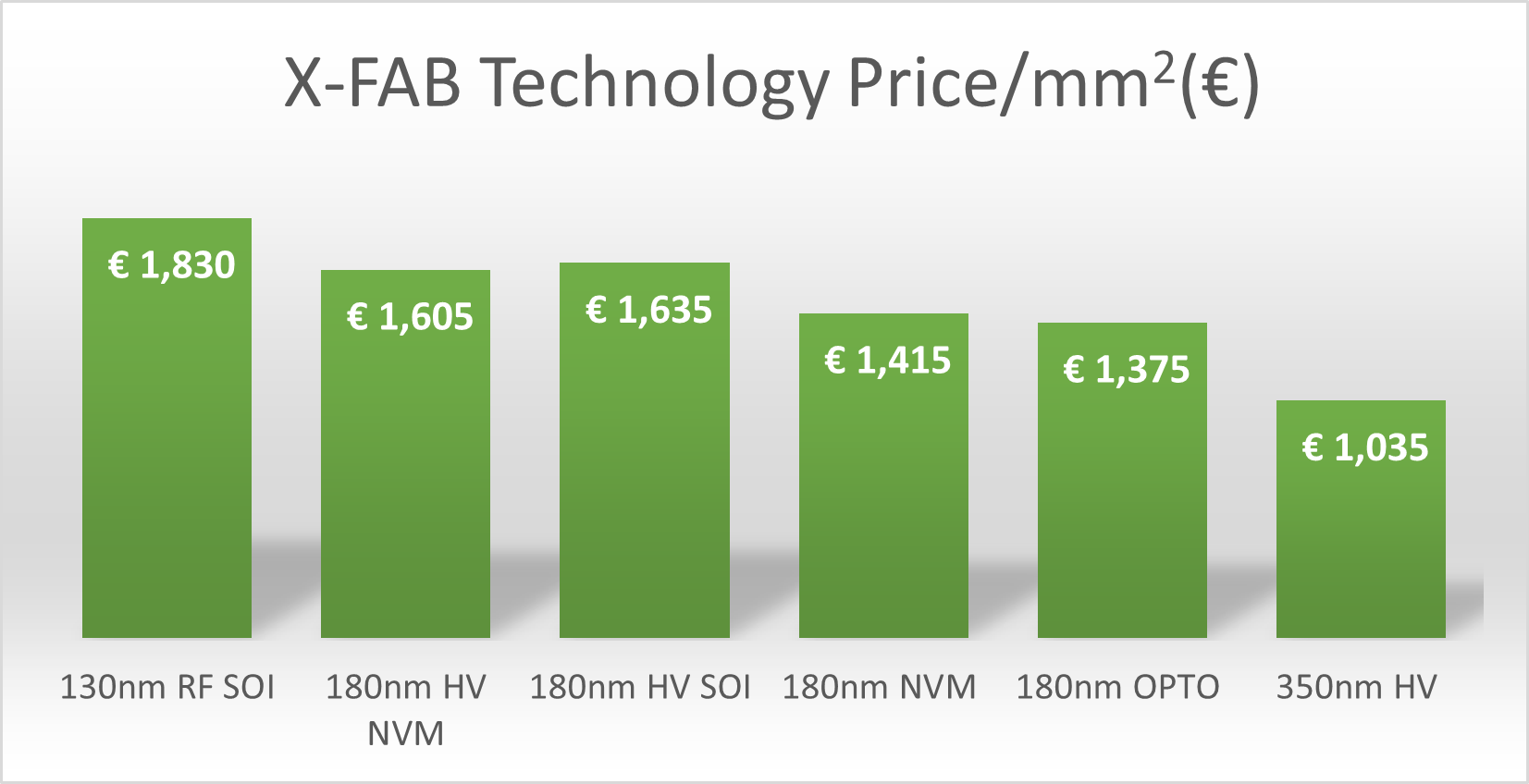}
\caption{X-FAB technology price per millimeter square [Source: Europractice \cite{Europractice}]}
\label{XFAB}
\end{figure}

\end{itemize}
\section{CMOS Technology Selection Strategy}
In order to get your product or research project to succeed, you have to understand the frequency of operation of your product. For example, 5G RF transmitter operates, ultrasonic systems operates in MHz while, EKG monitoring system for biomedical applications operates at  KHz. Consequently, 180 nm process is cost-efficient for biomedical application, however it won't be good enough in the GHz range.  

Thus, once we realize the application, the designer can start calculating the cost of production for each of the potential fabrication technologies that can operate at the required frequency.

\section{Price of Silicon Chip}

\begin{figure}
    \centering
    \includegraphics[width=0.5\textwidth]{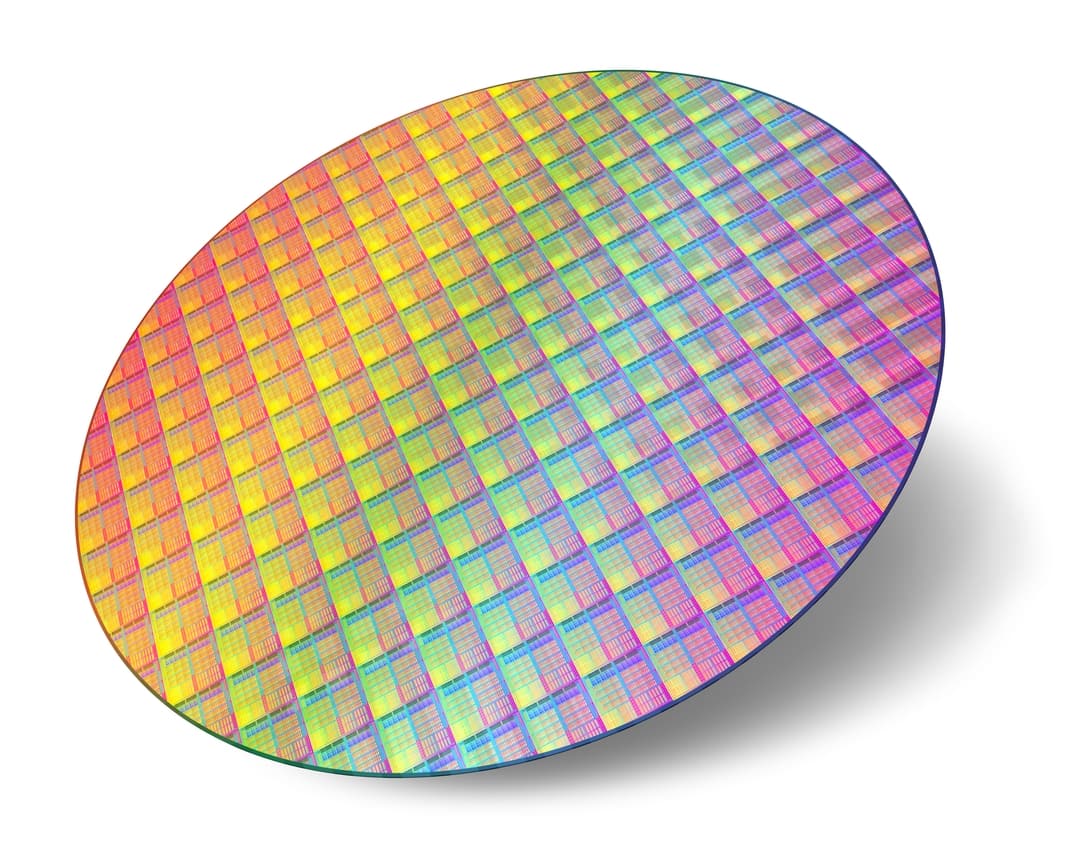}
    \caption{Silicon Wafer [Source: Wafer World \cite{wafer}]}
    \label{wafer}
\end{figure}

\begin{figure}
    \centering
    \includegraphics[width=0.7\textwidth]{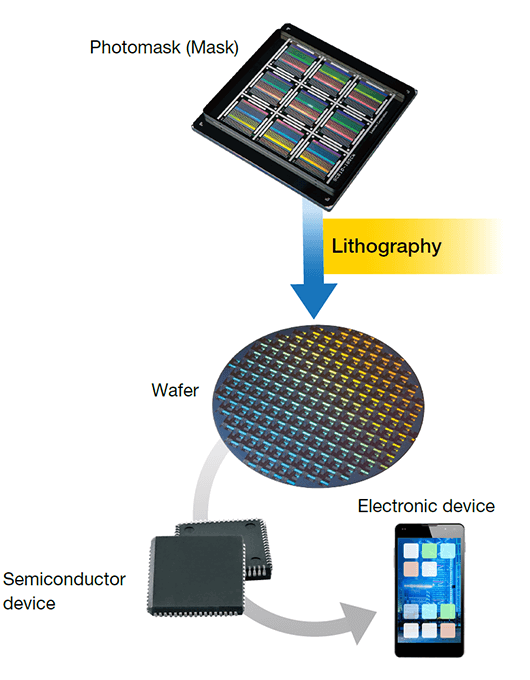}
    \caption{Electronics Fabrication Steps [Source: Lasertech \cite{lasertech}]}
    \label{electronic cycle}
\end{figure}

Silicon Wafers are used for fabricating integrated circuits. Figure \ref{wafer} depicts silicon wafer \cite{wafer}. After finishing the fabrication steps using photo mask and photo lithography, the wafer is diced into separate silicon chips, that gets packaged and assembled into a final product \cite{AMD}. Figure \ref{electronic cycle} summarizes the electronics fabrication stages \cite{lasertech}.

The majority of the cost for integrated circuits manufacturing goes to the fabrication of the photo masks for processing the wafers. The masks for 65nm technology cost roughly \$500,000 \cite{circuits_economics}. Additionally, the wafer cost is around \$2000-3000. Table \ref{technology_comparison} summarizes a comparison between the wafer diameters, photo mask and wafer cost for 350nm, 65 nm, and 14 nm technologies \cite{circuits_economics,14nmwafer}.

\begin{table}[ht]
    \centering
    \begin{tabular}{|c|c|c|c|}
         \hline
         Technology & Wafer Diameter & Mask Cost & Wafer Cost  \\
         \hline
         350 nm & 200 mm (8 inch) & \$30,000 & $<$\$500 \\
         \hline
         65 nm & 300 mm (12 inch) & \$500,000 & \$2000 \\
        \hline
         14 nm & 400 mm (16 inch) & $>$\$1,000,000 & $~$\$3900 \\
        \hline
    \end{tabular}
    \caption{Comparison of the wafer diameters, photo mask and wafer cost for 350nm, 65 nm, and 14nm technologies \cite{circuits_economics,14nmwafer}}
    \label{technology_comparison}
\end{table}

Consequently, the price of fabrication of one chip in 65nm technology is approximately \$502,000 in addition to the cost of the Electronic Design Automation (EDA) software, design engineers, packaging, and testing to verify the performance.

In case of mass production, the final cost of the silicon chip will depends on the number of chips per wafer, which depends on the area of the chip, utilization of the wafer, and yield.

\begin{figure}
    \centering
    \includegraphics[width=0.7\textwidth]{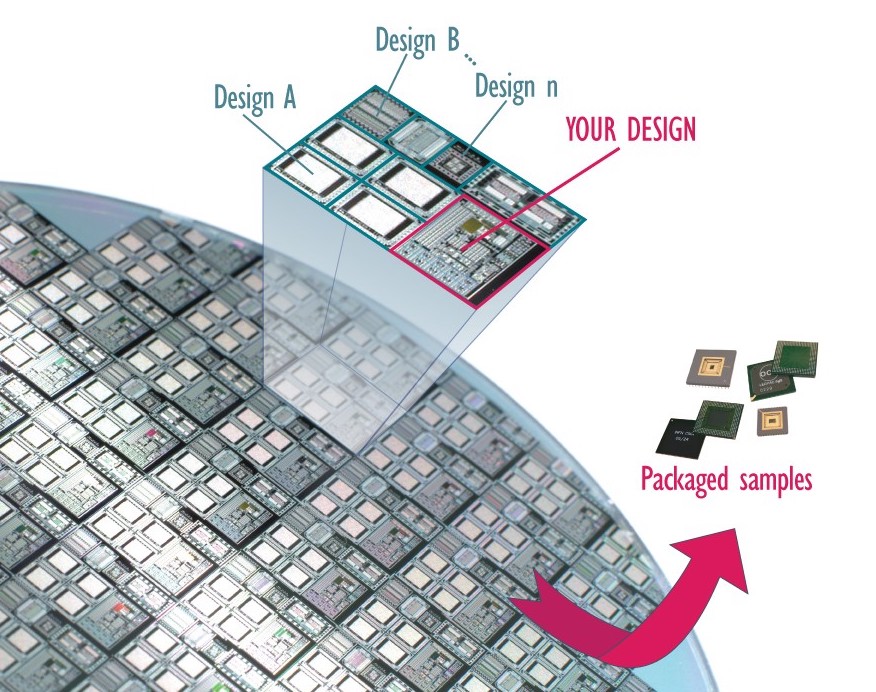}
    \caption{Multi-Project per Wafer [Source: Europractice \cite{Europractice_MPW}]}
    \label{mpw}
\end{figure}

If a number designers can share the same wafer to test their designs as shown in Figure \ref{mpw}, then the cost of the masks will be \$500,0000 divided by the number of the unique designs.

\bibliographystyle{IEEEtran}
\bibliography{References}

\begin{thebibliography}{1}
\providecommand{\url}[1]{#1}
\csname url@samestyle\endcsname
\providecommand{\newblock}{\relax}
\providecommand{\bibinfo}[2]{#2}
\providecommand{\BIBentrySTDinterwordspacing}{\spaceskip=0pt\relax}
\providecommand{\BIBentryALTinterwordstretchfactor}{4}
\providecommand{\BIBentryALTinterwordspacing}{\spaceskip=\fontdimen2\font plus
\BIBentryALTinterwordstretchfactor\fontdimen3\font minus
  \fontdimen4\font\relax}
\providecommand{\BIBforeignlanguage}[2]{{%
\expandafter\ifx\csname l@#1\endcsname\relax
\typeout{** WARNING: IEEEtran.bst: No hyphenation pattern has been}%
\typeout{** loaded for the language `#1'. Using the pattern for}%
\typeout{** the default language instead.}%
\else
\language=\csname l@#1\endcsname
\fi
#2}}
\providecommand{\BIBdecl}{\relax}
\BIBdecl

\bibitem{TSMC_comparison}
\BIBentryALTinterwordspacing
{Muse Semiconductor}. {TSMC MPW Shared Tapeouts}. [Online]. Available:
  \url{https://www.musesemi.com/shared-block-tapeout-pricing [Accessed: Aug.
  2022]}
\BIBentrySTDinterwordspacing

\bibitem{Europractice}
\BIBentryALTinterwordspacing
{Europractice}. {Schedules \& prices 2022}. [Online]. Available:
  \url{https://europractice-ic.com/schedules-prices-2022/ [Accessed: Aug.
  2022]}
\BIBentrySTDinterwordspacing

\bibitem{wafer}
\BIBentryALTinterwordspacing
{Wafer World}. {Silicon Wafer Manufacturers, Materials Used to Make
  Semiconductors}. [Online]. Available:
  \url{{https://www.waferworld.com/post/silicon-wafer-manufacturers-materials
  [Accessed: Aug. 2022]}}
\BIBentrySTDinterwordspacing

\bibitem{AMD}
\BIBentryALTinterwordspacing
AMD. {Introduction to Semiconductors: The Brains of Modern Electronics}.
  [Online]. Available:
  \url{{https://www.amd.com/en/technologies/introduction-to-semiconductors
  [Accessed: Aug. 2022]}}
\BIBentrySTDinterwordspacing

\bibitem{lasertech}
\BIBentryALTinterwordspacing
Lasertech. {Semiconductor-related Inspection Systems}. [Online]. Available:
  \url{{https://www.lasertec.co.jp/en/ir/individuals/semiconductor.html
  [Accessed: Aug. 2022]}}
\BIBentrySTDinterwordspacing

\bibitem{circuits_economics}
\BIBentryALTinterwordspacing
A.~Hajimiri. {Economics of Integrated Circuits, Yield, Pricing}. Caltech.
  [Online]. Available: \url{https://www.youtube.com/watch?v=pnCqLdMO3h8
  [Accessed: Aug. 2022]}
\BIBentrySTDinterwordspacing

\bibitem{14nmwafer}
\BIBentryALTinterwordspacing
{EETimes}. {FD SOI Benefits Rise at 14nm}. [Online]. Available:
  \url{{https://www.eetimes.com/fd-soi-benefits-rise-at-14nm/ [Accessed: Aug.
  2022]}}
\BIBentrySTDinterwordspacing

\bibitem{Europractice_MPW}
\BIBentryALTinterwordspacing
{Europractice}. {MPW FABRICATION}. [Online]. Available:
  \url{https://europractice-ic.com/services/fabrication/ [Accessed: Aug. 2022]}
\BIBentrySTDinterwordspacing

\end{thebibliography}

\end{document}